\documentclass{ws-mpla}

\begin{document}

\markboth{S. A. Alavi}
{Scattering in Noncommutative Quantum Mechanics}

\catchline{}{}{}{}{}

\title{SCATTERING IN NONCOMMUTATIVE QUANTUM MECHANICS}

\author{S. A. Alavi}

\address{Department of Physics, Tarbiat Moallem University of Sabzevar,\\
P.O.Box 397, Sabzevar, Iran\\
Sabzevar House of Physics, Javan-Sara, Asrar Avenue, Sabzevar, Iran.\\
alavi@sttu.ac.ir\\
alialavi@fastmail.us}

\maketitle

\pub{Received (Day Month Year)}{}

\begin{abstract}
We derive the correction due to noncommutativity of space on Born approximation, then the correction
for the case of Yukawa potential is explicitly calculated. The correction depends on the angle
of scattering. Using partial wave method it is shown that the conservation of the number of particles in 
elastic scattering is also valid in noncommutative spaces which means that the unitarity relation is 
held in noncommutative spaces. We also show that the noncommutativity of space has no effect on the optical theorem. Finally we study Gaussian function potential in noncommutative spaces 
which generates delta function potential as $\theta \rightarrow 0$.

\end{abstract}

\ccode{PACS Numbers: 03.65.-w, 02.20.a.}

\section{Introduction}

Recently there have been notable studies on the formulation and possible
experimental consequences of extensions of the standard (usual) quantum
mechanics in the noncommutative spaces.\cite{1}$^-$\cite{21}
The study on noncommutative  spaces is much important for 
understanding phenomena at short distances beyond the present
test of QED.

In field theories the noncommutativity is introduced by replacing the
standard product by the star product. For a manifold parameterized
by the coordinates $x_{i}$, the noncommutative relations can be
written as
\begin{equation}
[\hat{x}_{i},\hat{x}_{j}]=i\theta_{ij}\hspace{1.cm}[\hat{x}_{i},\hat{p}_{j}]=i\delta
_{ij}\hspace{1.cm} [\hat{p}_{i},\hat{p}_{j}]=0 ,
\end{equation}
where $\theta_{ij}=\frac{1}{2}\epsilon_{ijk}\theta_{k}$. NCQM is
formulated in the same way like the standard quantum mechanics SQM
(quantum mechanics in commutative spaces), that is in terms of the
same dynamical variables represented by operators in a Hilbert
space and a state vector that evolves according to the
Schroedinger equation
\begin{equation}
i\frac{d}{dt}|\psi>=H_{nc}|\psi>,
\end{equation}
we have taken into account $\hbar=1$. $H_{nc}\equiv H_{\theta}$
denotes the Hamiltonian for a given system in the noncommutative
space. In the literatures two
approaches have been considered  for constructing the NCQM:\\
a) $H_{\theta}=H$, so that the only difference between SQM and
NCQM is the presence of a nonzero $\theta$ in the commutator of
the position operators i.e. Eq.(1). \\
b) By deriving the Hamiltonian from the Moyal analog of the
standard Schroedinger equation:
\begin{equation}
i\frac{\partial}{\partial t}\psi(x,t)=H(p=\frac{1}{i}\nabla,x)\ast
\psi (x,t)\equiv H_{\theta}\psi(x,t),
\end{equation}
where $H(p,x)$ is the same Hamiltonian as in the standard theory,
and as we observe the $\theta$ - dependence enters now through the
star product.\cite{19} In [21], it is shown that these two
approaches lead to the same physical theory.\\
It has been shown that many physical quantities, for example the spectrum
of hydrogen atom,\cite{1} Lamb shift\cite{1} and
Berry's phase,\cite{17} receive corrections due to noncommutativity of
space for which these corrections depend on the noncommutativity
parameter. It is also shown that the noncommutativity of space has no
effect on algebras of observables of systems of identical particles and
on Pauli exclusion principle.\cite{{3},{18}}

Scattering theory is of broad interest and importance to all branches of
physics from particle to condensed matter physics. In this paper we study
some important aspects of scattering theory in the framework of
noncommutative quantum mechanics.

In SQM physicists usually use two methods to study 
scattering process:\\
a) Born approximation,\\
b) Partial wave method.\\
Let us start with the definition of star product:
\begin{equation}
(f\star g)(x)= exp \left( \frac{i}{2}\theta_{\mu\nu}\partial_{x_{\mu}}
\partial_{y_{\nu}} \right) f(x)g(y) \mid_{x=y}.
\end{equation}
Since the noncommutativity parameter, if it is non-zero, should be very small compared to the length 
scales of the system, one can consider the noncommutativity effects to the first order,
then we have
\begin{equation}
(f\star g)= f(x)g(x)+\frac{i}{2}\theta_{\mu\nu}\partial_{\mu}f\partial_{\nu}g+
{\mathcal{O}}(\theta^{2}).
\end{equation}
In what follows, we use this equation to study scattering in noncommutative quantum
mechanics.

\section{Born approximation in noncommutative spaces}

The matrix elements $M_{fi}$ are defined as follows
\begin{equation}
M_{fi}^{NC}=\langle \psi_{f}\mid V\mid \psi_{i}\rangle_{NC}=\int d^{3}r e^{-i\frac{\vec p_{f}.\vec r}{\hbar}} 
\star V(\vec r)\star e^{i\frac{\vec p_{i}.\vec r}{\hbar}}. 
\end{equation}
Using the definition of star product to the first order the
integrand can be expanded as\\

\begin{eqnarray}
&&e^{-i\frac{\vec p_{f}.\vec r}{\hbar}} V(\vec r) e^{i\frac{\vec p_{i}.\vec r}{\hbar}}+
\frac{i}{2}e^{-i\frac{\vec p_{f}.\vec r}{\hbar}} \theta_{\mu\nu}(\nabla V)_{\mu}(i\frac{\vec p_{\nu}^{i}}{\hbar})
e^{i\frac{\vec p_{i}.\vec r}{\hbar}}\nonumber\\
&+&
\frac{i}{2}e^{-i\frac{\vec p_{f}.\vec r}{\hbar}} \theta_{\mu\nu}(\nabla V)_{\nu}(-i\frac{\vec p_{\mu}^{f}}{\hbar})
e^{i\frac{\vec p_{i}.\vec r}{\hbar}}+\frac{i}{2}\theta_{\mu\nu} (-i\frac{\vec p_{\mu}^{f}}{\hbar}) (i\frac{\vec p_{\nu}^{i}}{\hbar})
V(\vec r)e^{i\frac{(\vec p_{i}-\vec p_{f}).\vec r}{\hbar}}.
\end{eqnarray}
Therefore we have\\
\begin{eqnarray}
\langle \psi_{f}\mid V\mid \psi_{i}\rangle_{NC}&=&
\langle \psi_{f}\mid V\mid \psi_{i}\rangle_{C}+
\int d^{3}r\frac{1}{2\hbar}\theta_{\mu\nu}(p^{i}_{\mu}+p^{f}_{\mu})(\nabla V)_{\nu}e^{i\frac{(\vec p_{i}-\vec p_{f}).\vec r}
{\hbar}}\nonumber\\
&+&\int d^{3}r\frac{i}{2\hbar^{2}}\theta_{\mu\nu} p^{f}_{\mu} p^{i}_{\nu} V(\vec r)
e^{i\frac{(\vec p_{i}-\vec p_{f}).\vec r}{\hbar}}.
\end{eqnarray}
The first term is the Born approximation in commutative space, the second and the third terms show the 
effects of noncommutativity of space.
Now let us derive these corrections, we have
\begin{equation}
\frac{i}{2\hbar^{2}}\theta_{\mu\nu}p_{\mu}^{f}p_{\nu}^{i}=\frac{i}{4\hbar^{2}}\epsilon_{\mu\nu\sigma}
p_{\mu}^{f}p_{\nu}^{i}\theta_{\sigma}=\frac{i}{4\hbar^{2}}(\vec p_{f}\times
\vec p_{i}).\vec\theta.
\end{equation}
If we take the component of $\vec\theta$ in the direction perpendicular to the plane of $\vec p_{f}$ and 
$\vec p_{i}$(z-axis) equal to $\theta_{3}=\theta$ and put the rest $\theta-$components to zero(which can be done by 
a rotation or a redifinition of coordinates), then we have
\begin{equation}
\frac{i}{4\hbar^{2}}(\vec p_{f}\times \vec{p}_{i}).\vec \theta
=\frac{i\theta}{4\hbar^{2}}p^{2}\sin\phi,
\end{equation}
where $\phi$ is the angle between $\vec p_{f}$ and $\vec p_{i}$.\\
In the same way we can calculate the contribution of the third term
in Eq.(8). Then for Yukawa potential
$V(r)=Z_{1}Z_{2}e^{2}\frac{e^{-\frac{r}{a}}}{r}$,
the differential cross section in a noncommutative space is
\begin{equation} 
(\frac{d\sigma}{d\Omega})_{NC}= \left( \frac{Z_{1}Z_{2}e^{2}}
{4E\sin^{2}(\frac{\phi}{2})+\frac{\hbar^{2}}{2ma^{2}}} \right)^{2}
\bigg[ 1+\frac{9\theta^{2} p^{4}}{16\hbar^{4}}\sin^{2}\phi \bigg].
\end{equation} 
We observe that the correction depends on the angle of
scattering $\phi$.

\section{Partial wave method in noncommutative spaces-Partial
wave expansion of scattering amplitude}

We consider a sphere of radius $r=R$. Inside the sphere there may be given a potential $V(r)$; outside 
the sphere the potential may vanish. A beam of particles described by a plane wave is scattered at this potential field. 
We shall compute the scattering amplitude by expansion into a series of partial waves.\\
In the domain $r<R$ the wave function can be written as
\begin{equation}   
u^{inside}(r,\theta^{\prime})= \sum_{\ell=0}^{\infty}i^{\ell}
\left( 2\ell +1 \right) \bigg[ \frac{1}{kr}\chi_{\ell}(kr) \bigg] \star
P_{\ell}(\cos\theta^{\prime}).
\end{equation}   
The boundary condition $\chi_{\ell}(0)=0$, ensures that $R_{\ell}(r)=\frac{\chi_{\ell}(r)}{r}$ be  finite at  $r=0$.\\
Outside the sphere $r=R$ we write
\begin{equation}  
u^{outside}(r,\theta^{\prime})= \sum_{\ell=0}^{\infty}i^{\ell}
\left( 2\ell +1 \right) \bigg\{ \frac{1}{kr} \bigg[ j_{\ell}(kr)+
\frac{1}{2}\alpha_{\ell}h_{\ell}^{(1)}(kr) \bigg] \bigg\} \star
P_{\ell}(\cos\theta^{\prime}).
\end{equation}   
Up to the first order in noncommutativity parameter $\theta$, we have
\begin{eqnarray}
u^{inside}(r,\theta^{\prime})&=&
\frac{1}{k}\sum_{\ell=0}^{\infty}i^{\ell} \left( 2\ell +1 \right)
\bigg\{ \frac{1}{r}\chi_{\ell}(kr)P_{\ell}
(\cos\theta^{\prime})\nonumber\\
&+&\frac{i}{2}\theta_{\mu\nu}\partial_{\mu} \left( \frac{1}{r}
\chi_{\ell}(r) \right) \partial_{\nu}P_{\ell}
(\cos\theta^{\prime}) \bigg\} + {\mathcal{O}}(\theta^{2})
\end{eqnarray}   
and
\begin{eqnarray}
u^{outside}(r,\theta^{\prime})&=&\frac{1}{k}\sum_{\ell=0}^{\infty}
i^{\ell} \left (2\ell +1 \right) \bigg\{ \bigg[ j_{\ell}(kr)+
\frac{1}{2}\alpha_{\ell}h_{\ell}^{(1)}(kr) \bigg] P_{\ell}
(\cos\theta^{\prime})\nonumber\\
&+&\frac{i}{2}\theta_{\mu\nu}\partial_{\mu} \bigg[\frac{1}{r}
(j_{\ell}(kr)+\frac{1}{2}\alpha_{\ell}h_{\ell}^{(1)}(kr)) \bigg]
\partial_{\nu}P_{\ell}(\cos\theta^{\prime}) \bigg\}+
{\mathcal{O}}(\theta^{2}).
\end{eqnarray}   
One can express the coefficients $\alpha_{\ell}$ by the
Logarithmic derivatives $L_{\ell}$
\begin{equation}   
L_{\ell}=\left (\frac{d\log \chi_{\ell}}{d\log r} \right)_{r=R}.
\end{equation}   
To do this, we use the continuity of both $\chi_{\ell}$ and $\frac{d\chi_{\ell}}{dr}$ when passing through the sphere 
$r=R$ and divide the second by the first one, then we get
\begin{eqnarray}   
L_{\ell}&=&x\frac{A+\frac{\theta}{2}B}{C+\frac{\theta}{2}D},\\
A&=&\frac{1}{x} \bigg[j^{\prime}_{\ell}(x)+\frac{1}{2}\alpha_{\ell}
h^{(1)\prime}_{\ell}(x) \bigg],\\
B&=&\frac{1}{x^{2}}j^{\prime}_{\ell}(x)+\frac{1}{2x^{2}}
\alpha_{\ell}h^{(1)\prime}_{\ell}(x)
-\frac{1}{x}j_{\ell}^{\prime\prime}(x)-\frac{1}{2x}
\alpha_{\ell}h^{(1)\prime\prime}_{\ell}(x),\\
C&=&\frac{1}{x} \bigg[j_{\ell}(x)+\frac{1}{2}
\alpha_{\ell}h^{(1)}_{\ell}(x) \bigg],\\
D&=&-\frac{1}{x^{2}}j_{\ell}(x)-\frac{1}{2x^{2}}
\alpha_{\ell}h^{(1)}_{\ell}(x)
+\frac{1}{x}j_{\ell}^{\prime}(x)+\frac{1}{2x}
\alpha_{\ell}h^{(1)\prime}_{\ell}(x),
\end{eqnarray}
where $x=kR$, and the prime denotes differentiation with respect to
the argument $kr$. Inversely we have
\begin{equation}
\alpha_{\ell}=2\frac{E+\frac{\theta}{2}F}{G+\frac{\theta}{2}H}
\end{equation}
where
\begin{eqnarray} 
E&=&-\frac{1}{x}L_{\ell}j_{\ell}(x)+j^{\prime}_{\ell}(x),\\
F&=&\frac{1}{x}j^{\prime}_{\ell}(x)+\frac{1}{x^{2}}
L_{\ell}j_{\ell}(x)-\frac{1}{x}L_{\ell}j^{\prime}_{\ell}(x)-
j^{\prime\prime}_{\ell}(x),\\
G&=&\frac{1}{x}L_{\ell}h^{(1)}_{\ell}(x)-h^{(1)\prime}_{\ell}(x),\\
H&=&-\frac{1}{x}h^{(1)\prime}_{\ell}(x)-\frac{1}{x^{2}}
L_{\ell}h^{(1)}_{\ell}(x)+\frac{1}{x}L_{\ell}
h^{(1)\prime}_{\ell}(x)+h^{(1)\prime\prime}_{\ell}(x),
\end{eqnarray}
where
$j_{\ell}(x)=\frac{1}{2}[h^{(1)}_{\ell}(x)+h^{(2)}_{\ell}(x)]$ and for real arguments, the spherical Hankel function of the second kind is  the complex 
conjugate of the first one.\\
In the limit of $\theta\rightarrow 0 $, $\alpha_{\ell}$ reduces
to its commutative counterpart\cite{22}
\begin{equation} 
\alpha_{\ell}= -2\frac{L_{\ell}j_{\ell}(x)-xj^{\prime}_{\ell}(x)}{L_{\ell}h^{(1)}_{\ell}(x)-
x h^{(1)\prime}_{\ell}(x)}.
\end{equation} 
After some calculations one can show that
\begin{equation} 
1+\alpha_{\ell}=-\frac{I+\frac{\theta}{2}K}{L+\frac{\theta}{2}M}
\end{equation} 
where
\begin{eqnarray} 
I&=&\frac{1}{x}L_{\ell}h^{(2)}_{\ell}(x)-h^{(2)\prime}_{\ell}(x),\\
K&=&-\frac{1}{x^{2}}L_{\ell}h^{(2)}_{\ell}(x)+\frac{1}{x}
L_{\ell}h^{(2)\prime}_{\ell}(x)-\frac{1}{x}h^{(2)\prime}_{\ell}(x)
+h^{(2)\prime\prime}_{\ell}(x),\\
L&=&\frac{1}{x}L_{\ell}h^{(1)}_{\ell}(x)-h^{(1)\prime}_{\ell}(x),\\
M&=&-\frac{1}{x^{2}}L_{\ell}h^{(1)}_{\ell}(x)+\frac{1}{x}L_{\ell}
h^{(1)\prime}_{\ell}(x)
-\frac{1}{x}h^{(1)\prime}_{\ell}(x)+h^{(1)\prime\prime}_{\ell}(x).
\end{eqnarray}
The numerator of $1+\alpha_{\ell}$ is complex conjugate of the
domeminator, so that
\begin{equation} 
\mid 1+\alpha_{\ell}\mid =1.
\end{equation}
This is an important result which shows the conservation of the number of particles in  elastic scattering in 
noncommutative spaces because the absolute squares of the amplitudes of ingoing and outgoing 
waves must be equal. This is the unitarity relation for the $\ell$th partial wave in a noncommutative space.\\
When $\theta\rightarrow 0 $ it reduces to 
\begin{equation} 
1+\alpha_{\ell}=-\frac{L_{\ell}h^{(2)}_{\ell}(x)-x h^{(2)\prime}_{\ell}(x)}{L_{\ell}h^{(1)}_{\ell}(x)-x h^{(1)\prime}_{\ell}(x)}
\end{equation}      
which again satisfies in Eq.(33).\cite{22}\\
Up to now, we have shown the unitarity to the first order in $\theta$.
One can show it for general case and to all
order in $\theta$, we have
\begin{equation}
\alpha_{\ell} +1=-\frac{L_{\ell}e^{\frac{\theta}{2}\frac{\partial}{\partial x}}[\frac{1}{x}
h_{\ell}^{(2)}(x)]-xe^{-\frac{\theta}{2}\frac{\partial}{\partial x}}[\frac{1}{x}h_{\ell}^{(2)\prime}(x)]}
{L_{\ell}e^{\frac{\theta}{2}\frac{\partial}{\partial x}}[\frac{1}{x}
h_{\ell}^{(1)}(x)]-xe^{-\frac{\theta}{2}\frac{\partial}{\partial x}}
[\frac{1}{x}h_{\ell}^{(1)\prime}(x)]},
\end{equation}
which satisfies in Eq.(33). To the zeroth and first order in $\theta$,
Eq.(35) reduces to Eq.(34) and Eq.(28), respectively.
The unitarity can be also shown without expanding in terms of Hankel
functions.\cite{4}\\
Finally one can easily shown that the noncommutativity of space
has no effect on the optical theorem and scattering
amplitude of two identical particles, because
\begin{equation}
\left( \frac{d\sigma}{d\Omega} \right)_{NC}
= \mid f(\phi)\mid^{2}=f(\phi )\star f^{*}(\phi),
\end{equation}
where
\begin{equation}
f(\phi)=\frac{1}{k}\sum_{\ell =0}^{\infty} \left( 2\ell
+1 \right) e^{i\delta_{\ell}(k)}\sin\delta_{\ell}(k)P_{\ell}(\cos\phi),
\end{equation} 
but $\phi$ is the angle between the ingoing and outgoing particle momenta. 
This angle variable has nothing to do with NC variables introduced in
Eq.(1) and therefore the star product
in Eq.(36), reduces to ordinary produt. The same argument is true for the case of scattering of two 
identical particles, we have\\
\begin{eqnarray}
\left( \frac {d\sigma}{d\Omega} \right)_{NC}
&=&\mid f(\phi)+f(\pi- \phi)\mid^{2}\nonumber\\
&=&f^{\star}(\phi)\star f(\phi)
+f^{\star}(\pi -\phi)\star f(\pi -\phi)+f^{\star}(\phi)\star
f(\pi -\phi)\nonumber\\
&+&f^{\star}(\pi -\phi)\star f(\phi),
\end{eqnarray} 
where again $\phi$ is the the angle between the ingoing and
outgoing particle momenta (two particles),
and therefore the noncommutativity of space has no effect on
this amplitude. This means that again the star product reduces
to ordinary product.

\section{Delta function potential}
Let us study scattering from delta function potential $V(\vec{r})=B\delta (\vec{r})$. The differential cross section 
in commutative case is isotropic
\begin{equation}  
\frac{d\sigma}{d\Omega}=\frac{\mu^{2}B^{2}}{4\pi^{2}\hbar^{4}}
\end{equation}   
where $\mu$ is the reduced mass. But we are not quite sure if we understand what a delta function potential 
in a NC setup would mean, because $[\hat{x}_{i},\hat{x}_{j}]=i\theta_{ij}$ implies that we can not localize
anything with infinite precision . Probably a good regulator for the delta function potential is a Gaussian 
function like $V(r)=\lambda\exp[-(\frac{r}{R})^{2}]$,  
which generates delta function as we send $\theta$ to zero, if we take $R=\frac{1}{\lambda}=\theta$. So we consider the Gaussian function in a NC space. 
Using Eq.(8), one can calculate the corrections due to space noncommutativity on the amplitude. Then the 
cross section will have the following form
\begin{equation} 
\left( \frac{d\sigma}{d\Omega} \right)_{NC}
=\frac{2\pi m^{2}\lambda^{2}}{\hbar ^{4}}R^{6}\exp
\bigg( - \bigg( \frac{8mE}{\hbar^{2}} \bigg) R^{2}\sin^{2}\frac{\phi}{2} \bigg)
\bigg[ 1+\frac{9\theta^{2} p^{4}}{16\hbar^{4}}\sin^{2}\phi \bigg].
\end{equation}   
In conclusion some important aspects of scatterring in noncommutative spaces has been studied and 
these results are interesting to further understand the basic physics in noncommutative
spaces.

\section*{Acknowledgments}
I would like to thank P. Pre\v{s}najder (Comenius University, Slovakia)
and M. M. Sheikh-Jabbari (Stanford University)
for their  careful reading of the manuscript and for their valuable
comments.

\end{document}